\documentclass[10pt,aps,pra,twocolumn,superscriptaddress,amsmath,amssymb,floatfix,nofootinbib,longbibliography]{revtex4-2}

\usepackage{graphicx}
\usepackage[utf8]{inputenc}
\usepackage{physics}
\usepackage{bbold}

\newcommand{\mt}[1]{\text{#1}}

\newcommand{\mkPM}{\raisebox{0ex}{\includegraphics[width=1.2ex,height=1.2ex]{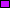}}}
\newcommand{\mkFM}{\raisebox{0ex}{\includegraphics[width=1.3ex,height=1.3ex]{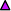}}}
\newcommand{\mkFive}{\raisebox{-0.05ex}{\includegraphics[width=1.2ex,height=1.2ex]{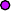}}}
\newcommand{\mkFFM}{\raisebox{-0.08ex}{\includegraphics[width=1.5ex,height=1.5ex]{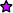}}}

\newcommand{\mkPMal}{\raisebox{-0.2ex}{\includegraphics[width=1.7ex,height=1.7ex]{mkRect.pdf}}}
\newcommand{\mkFMal}{\raisebox{-0.24ex}{\includegraphics[width=1.8ex,height=1.8ex]{mkTri.pdf}}}
\newcommand{\mkFiveal}{\raisebox{-0.23ex}{\includegraphics[width=1.8ex,height=1.8ex]{mkCirc.pdf}}}
\newcommand{\mkFFMal}{\hspace*{0.05em}\raisebox{-0.2ex}{\includegraphics[width=2.0ex,height=2.0ex]{mkStar.pdf}}\hspace*{-0.05em}}

\usepackage[colorlinks=true,linkcolor=blue,citecolor=blue,urlcolor=blue]{hyperref}
\begin{document}

\title{Ergodicity breaking for data classification in a dissipative Ising model with competing two- and four-body interactions}

\author{Simon Kochsiek}
\email{simon.kochsiek@uni-tuebingen.de}
\affiliation{Institut f\"ur Theoretische Physik, Universit\"at T\"ubingen and Center for Integrated Quantum Science and Technology, Auf der Morgenstelle 14, 72076 T\"ubingen, Germany}
\author{Mario Boneberg}
\affiliation{Institut f\"ur Theoretische Physik, Universit\"at T\"ubingen and Center for Integrated Quantum Science and Technology, Auf der Morgenstelle 14, 72076 T\"ubingen, Germany}
\author{Igor Lesanovsky}
\affiliation{Institut f\"ur Theoretische Physik, Universit\"at T\"ubingen and Center for Integrated Quantum Science and Technology, Auf der Morgenstelle 14, 72076 T\"ubingen, Germany}

\date{\today}
\begin{abstract}
    We explore the close relationship between the processing of data and the dynamics of interacting many-body systems. For example, certain classes of quantum neural networks can be mapped onto dissipative quantum Ising models whose evolution is described by a Lindblad master equation. In these architectures, emergent nonlinearities lead to non-ergodic dynamics under which initial states located in different basins of attraction evolve towards distinguishable stationary states. This mechanism underlies the classification of quantum data. Motivated by this connection, we consider an Ising model with competing two- and four-body interactions. Using a combination of mean-field calculations and exact small-scale numerics, we show that the stationary state manifold can be designed to enable multi-class classification. We also investigate the structure of the ensuing basins of attraction, thereby showing that these systems allow for the classification of quantum states with nonlinear decision boundaries.
\end{abstract}

\maketitle

\section{Introduction}
\label{sec:intro}
Classification, i.e. the assignment of inputs to a discrete set of labels, is among the most frequent tasks in machine learning \cite{bishop2006,murphy2012}.
It condenses complex data into a few simple categories, prominent examples being handwritten digit recognition~\cite{LeCun1998,nielsen2015} or inference of medical diagnoses from laboratory data~\cite{Rajkomar2019, Guncar2018, Ning2025}. Such tasks can be routinely handled by
neural networks. These pass input data through layers of computational units (neurons). Each neuron connects to a number of neurons in the previous layer, weighs and integrates their outputs and nonlinearly transforms them by means of an activation function.
Architectural features such as neuron connectivity and the nonlinear activation function have been demonstrated to have a great impact on classification, determining for instance the complexity of decision boundaries \cite{LeCun2015, Goodfellow2016, hornik1989}.

Significant research effort is currently dedicated to understanding~\cite{Biamonte2017, Dunjko2018, Cerezo2021, Cerezo2022, Wang2024, Larocca2025} and realizing~\cite{Havlicek2019, Herrmann2022} classification tasks on quantum devices.
One motivation for this is that quantum data can be processed natively~\cite{Huang2022, Cerezo2022}, i.e. without the potentially large overhead associated with representing quantum states on classical hardware. Indeed, general quantum states can feature a number of probability amplitudes that scales exponentially with the number of qubits~\cite{Georgescu2014, Preskill2018}. Converting them into a classical representation by means of a so-called state tomography~\cite{ODonnell2016, Haah2017} often requires an exponential number of measurements.

\begin{figure}[tbp]
    \centering
    \includegraphics[width=\linewidth]{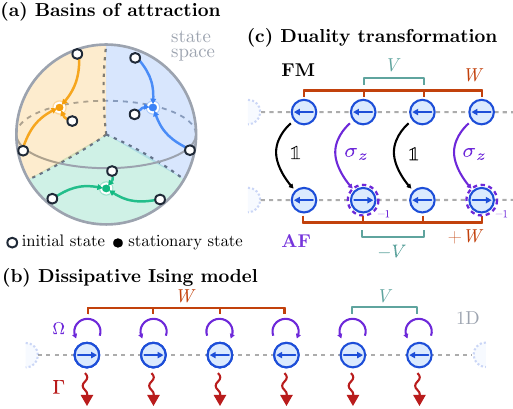}
    \caption{\textbf{Classification of quantum states by dissipative dynamics in a spin chain with two- and four-body interactions.}
        \textbf{(a)} Schematic representation of the state space. Initial states (white circles) evolve towards one of the stationary states (filled circles). All initial states in the same basin of attraction (colored region) are assigned to the same class. Dotted lines mark the decision boundaries.
        \textbf{(b)} Chain with spins oriented along \(x\)-axis (blue arrows) featuring competing processes: transverse field \(\Omega\) (purple arrows), decay at rate \(\Gamma\) (wavy arrows), nearest-neighbor coupling \(V\) and four-body coupling \(W\) on consecutive sites (see main text Sec.~\ref{sec:model}).
        \textbf{(c)} Duality transformation: the \(x\) and \(y\) components of every odd spin (dashed circles) are flipped, reversing the sign of \(V\), while other parameters are unchanged (Sec.~\ref{sec:duality}).}
    \label{fig:concept}
\end{figure}

The classification of quantum states has already been addressed extensively in the quantum machine learning literature in the past years, for instance by employing quantum versions of neural networks~\cite{Cong2019, Beer2020, Abbas2021, Meyer2023}. However, unlike classical neural networks, these models process data according to quantum mechanics, so states can only be transformed linearly by quantum operations. How nonlinearities can be introduced and leveraged in quantum neural network computation in a meaningful way remains a question of active research~\cite{Schuld2014, Cao2017,Holmes2023,Argyle2026}.

Recently, it has been shown that \textit{dissipative} quantum neural networks may exhibit effective nonlinearities that emerge with system size from a local, unitary gate sequence. Refs.~\cite{BonebergNonlinear2025,Boneberg2025} demonstrate an architecture that can be equivalently expressed as a Markovian discrete-time dissipative dynamics on a spin chain, generated by a quantum channel. In continuous time, such dynamics may feature dissipative phase transitions and related dynamical phenomena such as metastability~\cite{marcuzzi_universal_2014, Macieszczak2016, Minganti2018}.

These phenomena can also manifest at the level of the discrete-time dynamics of quantum neural networks, which allows to establish a close connection between the information processing on quantum states and interacting dissipative quantum spin models.
This idea broadly links to that of the paradigmatic Hopfield model~\cite{Hopfield1982,amit1985a,amit1985b,amit1987}, which is a classical spin system that mimics the behavior of an associative memory.
Its stationary states, the so-called patterns, are set by the choice of the spin couplings.
The dynamics evolves every initial state towards the closest stationary state, thereby retrieving the nearest (or most similar) pattern.
This non-ergodic behavior can be interpreted as classification: the initial state is the input, the stationary state it reaches is the output, and all initial states in the basin of attraction of the same stationary state are assigned the same class, as sketched in Fig.~\ref{fig:concept}(a).

In this work, we investigate a dissipative quantum Ising model with competing two- and four-body couplings~[Fig.~\ref{fig:concept}(b)]. Motivated by the possibility of mapping interacting spin systems onto quantum neural networks, we explore how the coupling range and effective nonlinearities give rise to ergodicity-breaking behavior. For this, we conduct a study that is based on a combination of mean-field calculations and exact numerics and study the structure of the stationary state manifold. We find that the four-body term dramatically increases the complexity of the stationary phase diagram and affects the number of different stationary states reached by the dynamics.
These results may inform the design of quantum neural networks, as they indicate how increasing the gate connectivity can enhance the expressive power of these architectures to establish multiclass classification capabilities.

The paper is organized as follows. Section~\ref{sec:model} introduces the model.
In Section~\ref{sec:symmetries} we investigate its symmetry properties, which inform the choice of order parameters that are used to study the stationary state manifold.
Section~\ref{sec:mft} develops a mean-field theory for the dynamical behavior of these order parameters.
Therein, we compute the stationary solutions (fixed points) of the corresponding mean-field equations of motion and analyze their stability.
Moreover, we show how manifest and emergent symmetries shape the structure of the stationary state manifold.
In Sec.~\ref{sec:dynamicsbasins} we investigate the approach towards the fixed points and map out their associated basins of attraction.
Section~\ref{sec:ed_method} is dedicated to the analysis of finite systems in which ergodicity cannot be strictly broken. Here we discuss which features of the mean-field study persist and how they manifest.
We discuss the implications of our results for the classification of quantum data in Sec.~\ref{sec:conclusion}.

\section{Model}
\label{sec:model}
In this section we introduce the dissipative Ising model considered in this work.
As illustrated in Fig.~\ref{fig:concept}(b), it is defined on a chain of spin-\({1}/{2}\) particles, with an even number of spins \(N\) and periodic boundary conditions. The density matrix \(\rho\), describing the state of the system, evolves under the Lindblad master equation (\(\hbar = 1\))
\begin{equation}
    \dot{\rho} = \mathcal{L}[\rho]
    = -i\commutator{H}{\rho}
    + \sum_{k=1}^{N}\left( L_k\rho L_k^{\dagger} - \frac{1}{2}\anticommutator{L_k^{\dagger}L_k}{\rho} \right),
    \label{eq:masterequation}
\end{equation}
with the \textit{Liouvillian} \(\mathcal{L}\). The first term generates coherent evolution, while the sum, known as the \textit{dissipator}, describes incoherent processes.
The Hamiltonian
\begin{equation}
    H = \Omega\sum_{k=1}^{N}\sigma_k^z
    + V\sum_{k=1}^{N}\sigma_k^x\sigma_{k+1}^x
    + W\sum_{k=1}^{N}\sigma_k^x\sigma_{k+1}^x\sigma_{k+2}^x\sigma_{k+3}^x
    \label{eq:hamiltonian}
\end{equation}
contains three contributions, that are written in terms of the Pauli operators \(\sigma_k^\alpha\) (\(\alpha=x,y,z\)) on site \(k\). Since we consider periodic boundary conditions, we have \(\sigma_{N+k}^x = \sigma_{k}^x\). The first term describes a transverse field (in \(z\)-direction) of strength \(\Omega\). The second term represents a nearest-neighbor Ising interaction of strength \(V\) along the \(x\)-direction, and the third term describes a four-body interaction at strength \(W\). The jump operators, which govern the dissipative processes in the Liouvillian \(\mathcal{L}\), are given by
\begin{equation}
    L_k = \sqrt{\Gamma}\,\sigma_k^-,\qquad \sigma_k^- = \big(\sigma_k^x - i\sigma_k^y\big) / 2,
    \label{eq:jumpoperators}
\end{equation}
and describe the decay of each spin at rate \(\Gamma\). For \(W = 0\) the model reduces to the dissipative Ising model discussed in Refs.~\cite{Weimer2015, Overbeck2017, Boneberg2025} and multi-site interactions in this context have been discussed in Ref.~\cite{Wang2021}.

The dynamics governed by Eq.~\eqref{eq:masterequation} features several competing processes. Taken individually, the transverse field acts as a coherent drive and rotates each spin about the \(z\)-axis, while the evolution under the interaction terms induces spin rotations about the \(x\)-axis with an angle that is set by the orientation of either one or three neighboring spins. The decay, in turn, forces the spins towards the state \(\ket{\downarrow}\). The interplay of these processes leads to a non-trivial manifold of stationary states and complex relaxation towards it. To analyze this systematically, it is convenient to eliminate one parameter. Given that we are mainly interested in understanding the impact of four-body interactions, we keep \(W\) finite and positive and use it as the unit of energy. This entails the rescaling
\begin{equation}
    \Omega/W \to \Omega,\quad V/W \to V,\quad \Gamma/W \to \Gamma,\quad tW \to t,
\end{equation}
so that \(W = 1\) and \(\Omega\), \(V\), \(\Gamma\) and \(t\) are dimensionless.

\section{Symmetries \& order parameters}
\label{sec:symmetries}
Before considering the dynamics of the model, we can already constrain the possible structure of the stationary manifold through the symmetries of the Liouvillian governing Eq.~\eqref{eq:masterequation}. In the following, we show that the model possesses a weak \(\mathbb{Z}_2\) symmetry which, in certain parameter regimes, makes the stationary states appear in pairs. Moreover, there exists a duality transformation, which flips the sign of the nearest-neighbor interaction \(V\) and leaves all other parameters unchanged. This transformation establishes a relation between stationary states in different parameter regions. We conclude this section by discussing how these findings inform the choice of order parameters that we use for studying the stationary state manifold and the evolution of the system towards it.

\subsection{Global parity}
\label{sec:parity}
The Liouvillian \(\mathcal{L}\) in Eq.~\eqref{eq:masterequation} commutes with the parity superoperator \(\mathcal{P}\), defined by
\begin{equation}
    \mathcal{P}[\rho] = P\rho P,\qquad P = \prod_{k=1}^{N}\sigma_k^z .
    \label{eq:parity}
\end{equation}
This operation performs the mapping \(\sigma_k^{x,y}\to-\sigma_k^{x,y}\) on every site \(k\) and leaves the operators \(\sigma_k^{z}\) invariant. Since the Hamiltonian \eqref{eq:hamiltonian} is composed of terms that either contain \(\sigma_k^{z}\) or an even number of \(\sigma_k^{x}\) it is also invariant under this parity transformation. The jump operators, on the other hand, change the sign, \(\mathcal{P}[\sigma_k^-] = -\sigma_k^-\). But since these enter the dissipator quadratically, this sign change cancels such that \(\commutator{\mathcal{L}}{\mathcal{P}} = 0\). Such a symmetry, which holds for the Liouvillian as a whole but not necessarily on the level of the individual jump operators, is called a \emph{weak} symmetry~\cite{Buca2012, Albert2014, Minganti2018}. Unlike a strong symmetry, which also leaves the jump operators invariant under the symmetry transformation, it has no associated conserved quantity and does not imply that the Liouvillian is reducible and possesses more than one stationary state.
Nevertheless, the weak symmetry has an impact on the structure of the stationary state manifold: if \(\rho_\mt{ss}\) is a stationary state, so is \(\mathcal{P}[\rho_\mt{ss}]\). This means that either the stationary state is unique, in which case it is symmetric, \(\mathcal{P}[\rho_{\mathrm{ss}}] = \rho_{\mathrm{ss}}\), and the \(x\) and \(y\) components of all spin operators (Pauli matrices) vanish on average. Or the symmetry is broken spontaneously (possible only in the thermodynamic limit~\cite{Buca2012, lieu2020}) and stationary states come in pairs, \(\rho^+_\text{ss}\) and \(\rho^-_\text{ss} = \mathcal{P}[\rho^+_\text{ss}]\).

\subsection{Duality transformation}
\label{sec:duality}
In the following we discuss a transformation which in general does not leave the master equation \eqref{eq:masterequation} invariant, but maps the model onto itself with a different set of parameters. The transformation, which is illustrated in Fig.~\ref{fig:concept}(c), consists of a \(\pi\)-rotation of the spins about their \(z\)-axis on every odd site. This is implemented by the unitary operation
\begin{equation}
    U = \prod_{k\,\mathrm{odd}}\sigma_k^z,
    \qquad
    U\sigma_k^{x,y}U^\dagger = (-1)^k\,\sigma_k^{x,y}.
    \label{eq:sublatticeparity}
\end{equation}
In Hamiltonian \eqref{eq:hamiltonian} this reverses the sign of the two-body interactions: \(V\rightarrow -V\). However, the four-body interaction remains invariant. The jump operators merely acquire a phase, \(\sigma_k^-\to(-1)^k\sigma_k^-\), which cancels when constructing the dissipator. Using the unitary superoperator \(\mathcal{U}[\rho]=U\rho U^\dagger\) we therefore find
\begin{equation}
    \mathcal{U}\,\mathcal{L}(\Omega,V,\Gamma)\,\mathcal{U}^{-1}
    = \mathcal{L}(\Omega,-V,\Gamma).
    \label{eq:duality}
\end{equation}
For \(V=0\) this becomes a weak symmetry, but for \(V \neq 0\), it is an operation that connects two different models. It implies that eigenvalues of the Liouvillian are unchanged when \(V\rightarrow-V\), and that the eigenmodes are mapped onto each other via \(\rho_i\mapsto U\rho_i U^\dagger\). In particular, for the stationary state one finds
\begin{equation}
    \rho_{\mathrm{ss}}(\Omega,-V,\Gamma) = U\,\rho_{\mathrm{ss}}(\Omega,V,\Gamma)\,U^\dagger,
    \label{eq:ssduality}
\end{equation}
i.e., a stationary state at interaction strength \(V\) has a counterpart at \(-V\), which is obtained by flipping the \(x\) and \(y\) components of the spins on every odd site.

\subsection{Order parameters}
\label{sec:orderparameters}
The behavior of the model under these two transformations allows us to identify suitable order parameters.
A reasonable assumption is that the weak \(\mathbb{Z}_2\) symmetry associated with the parity superoperator \(\mathcal{P}\) can be broken spontaneously.
In the symmetry-broken phase, there exists for each stationary state \(\rho_{\mathrm{ss}}^-\) an associated symmetry-transformed stationary state \(\rho_{\mathrm{ss}}^+ = \mathcal{P}[\rho_{\mathrm{ss}}^-]\). Stationary states related by such a symmetry transformation differ in the sign of the local expectation values \(\ev{\sigma_k^{x,y}}\) while the expectation value \(\ev{\sigma_k^z}\) remains unchanged. A suitable order parameter therefore should be sensitive to the spin components perpendicular to \(z\). This motivates the definition of the \textit{magnetization} operator
\begin{equation}\label{eq:uniformmagnetization}
    \hat m_x = \frac{1}{N}\sum_{k=1}^{N}\sigma_k^x .
\end{equation}
A nonzero expectation value of this operator signals ferromagnetic (FM), i.e. uniform ordering of the spins and allows one to distinguish the different symmetry-related stationary states. On the other hand, the duality transformation \(\mathcal{U}\) relates the stationary states of parameter regimes \(\pm V\) by means of a \(\pi\)-rotation of the spins about their \(z\)-axis on every odd site. Thus, if for \(+V\) the stationary manifold can be characterized by \(\ev{\hat m_x}_{\mathrm{ss}}(V)= \tr(\hat m_x \rho_{\mathrm{ss}}(\Omega,V,\Gamma))\), then for \(-V\) it is the  \textit{staggered magnetization} operator
\begin{equation}\label{eq:staggeredmagnetization}
    \hat n_x = \frac{1}{N}\sum_{k=1}^{N}(-1)^k\,\sigma_k^x
\end{equation}
that characterizes the stationary state manifold, since \(\hat n_x = U\hat m_x U^\dagger\). Therefore it follows
\[
    \ev{\hat n_x}_{\mathrm{ss}}(-V)= \tr(\hat n_x \rho_{\mathrm{ss}}(\Omega,-V,\Gamma)) = \ev{\hat m_x}_{\mathrm{ss}}(V).
\]
A nonzero expectation value of \(\hat n_x\) signals antiferromagnetic (AF) or staggered ordering. In contrast a paramagnetic (PM) stationary state does not break the symmetry and has \(\expval{\hat m_x}_{\mathrm{ss}} = \expval{\hat n_x}_{\mathrm{ss}} = 0\).

\section{Mean-field theory}
\label{sec:mft}
In order to obtain a first understanding of the many-body dynamics and the stationary states we perform in the following a mean-field analysis. To this end, we derive the equations of motion for the order parameters starting from the adjoint master equation.
Subsequently, we compute the stationary solutions and study their stability.
In this context we also discuss the impact of the symmetries on the structure of the stationary state manifold, which leads us to the mean-field phase diagram.
Finally, we investigate the structure of the basins of attraction, i.e., we analyze which regions of the mean-field state space are drawn to which stationary states. From the perspective of data processing, this hints at the possibility to construct complex, i.e., nonlinear decision boundaries for the classification of quantum states.

\subsection{Mean-field equations of motion}
\label{sec:mf_sublattice}
In the Heisenberg picture, an operator \(\hat{O}\) evolves under the adjoint master equation
\begin{equation}
    \begin{split}
        \partial_t \hat{O} & = \mathcal{L}^\dagger[\hat{O}] \\
                           & = i\commutator*{H}{\hat{O}}
        + \sum_k \left( L_k^\dagger \hat{O} L_k
        - \tfrac{1}{2}\anticommutator*{L_k^\dagger L_k}{\hat{O}} \right).
    \end{split}
    \label{eq:adjoint}
\end{equation}
Expectation values of operators, such as the magnetization \({\ev*{\hat{m}_x} = N^{-1}\sum_k \tr(\sigma^x_k\rho)}\), hence follow the differential equation \(\partial_t\ev*{\sigma^x_k} = \ev*{\mathcal{L}^\dagger[\sigma^x_k]}\). In general, however, this results in an infinite hierarchy of coupled equations of expectation values and higher-order correlation functions. To enforce its closure at the level of single-site expectation values, we assume that for all times the density matrix can be written as a product state of the form
\begin{equation}\label{eq:meanfieldstate}
    \rho = \rho_A \otimes \rho_B \otimes \rho_A \otimes \rho_B \otimes \cdots .
\end{equation}
This state depends on the two sublattice density matrices \(\rho_{A},\,\rho_B\), which are parametrized using the sublattice Bloch vectors with entries \(\nu_{A/B} = \ev{\sigma^\nu}_{A/B}\) (with \(\nu=x,y,z\)):
\begin{equation}
    \rho_{A/B} = \frac{1}{2}\left( \mathbb{1} + \sum_{\nu=x,y,z} \nu_{A/B}\, \sigma^\nu \right).\label{eq:blochvectors}
\end{equation}
The mean-field equations of motion can then be recast into a set of nonlinear coupled differential equations for the overall six components of the two Bloch vectors \({\vec m_A = (x_A, y_A, z_A)}\) and \({\vec m_B = (x_B, y_B, z_B)}\), which characterize the mean-field state of each sublattice (see Appendix~\ref{app:mf}).
They obey
\begin{align}
    \dot x_{A/B} & = -2\Omega\, y_{A/B} - \frac{\Gamma}{2}\, x_{A/B},\label{eq:sublatticeX}                      \\
    \dot y_{A/B} & = 2\Omega\, x_{A/B} - h_{AB/BA}\, z_{A/B} - \frac{\Gamma}{2}\, y_{A/B},\label{eq:sublatticeY} \\
    \dot z_{A/B} & = h_{AB/BA}\, y_{A/B} - \Gamma\left(z_{A/B} + 1\right).\label{eq:sublatticeZ}
\end{align}
The coupling between the sublattices is mediated through the \textit{effective fields}
\begin{equation}
    \begin{aligned}
        h_{AB} & = 4V\, x_B + 8 x_A x_B^2, \\
        h_{BA} & = 4V\, x_A + 8 x_B x_A^2.
    \end{aligned}
    \label{eq:fields}
\end{equation}
For a translation invariant state the fields are symmetric under exchange of the sublattices, i.e. \(h_{AB} = h_{BA}\), and the equations for the dynamics on the sublattices A and B coincide (see Appendix~\ref{app:mf}).
Note that the order parameters, i.e. the operators in Eqs.~\eqref{eq:uniformmagnetization} and~\eqref{eq:staggeredmagnetization} evaluated on the state \eqref{eq:meanfieldstate}, are related to the sublattice Bloch vectors \(\vec m_A\) and \(\vec m_B\) by
\begin{equation}
    \begin{aligned}
        m_x & = \expval{\hat m_x} = \frac{x_A + x_B}{2}, \\
        n_x & = \expval{\hat n_x} = \frac{x_A - x_B}{2}.
    \end{aligned}
    \label{eq:mf_orderparameters}
\end{equation}

\begin{figure*}[tbp]
    \makebox[\textwidth][r]{
        \includegraphics[width=1.03\linewidth]{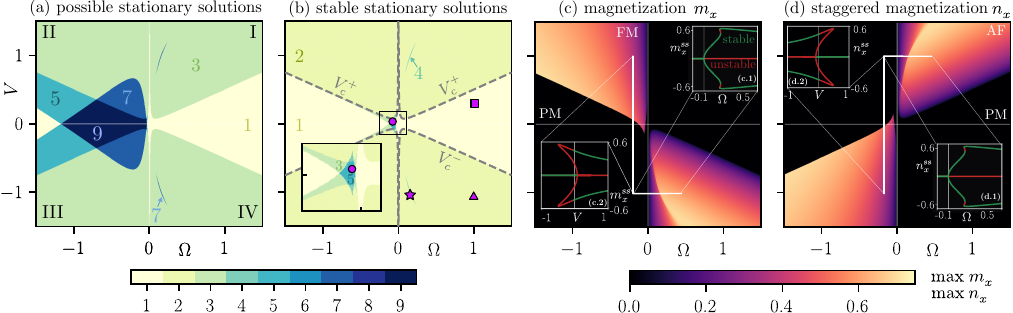}
    }
    \caption{\textbf{Mean-field phase diagram.} The number of solutions and corresponding order parameters plotted against the system parameters \((\Omega, V)\) in the mean-field approximation (see Sec.~\ref{sec:mft}). Results are calculated numerically using the parameter values \(W=1\) and \(\Gamma= 0.3\).
        \textbf{(a)} Number of physical stationary solutions of Eqs.~\eqref{eq:sublatticeX}-\eqref{eq:sublatticeZ}. The symmetry of the plot around \(V=0\) is a result of the duality transformation (see Sec.~\ref{sec:duality}).
        \textbf{(b)} Number of \emph{stable} stationary solutions of Eqs.~\eqref{eq:sublatticeX}-\eqref{eq:sublatticeZ} obtained by linear stability analysis (see Sec.~\ref{sec:mf_stability}). The paramagnetic solution
            [PM, see Eq.~\eqref{eq:paramagnetic_state}] is stable in between the dashed lines for \(|V| < V_\mathrm{c}\) [see Eq.~\eqref{eq:pm_stability}]. The inset is a zoomed-in region with \(\Omega \in [-0.25, 0.1], V\in [-0.1,0.1]\), where a maximum of 5 solutions can be stable simultaneously. Four representative parameter points \((\Omega , V)\) are chosen for further analysis: a paramagnetic point \mkPM~~\(= (1.0, 0.2)\), a ferromagnetic (FM) point \mkFM~~\(=(1,-1)\), a multistable point with four solutions \mkFFM~~\(=(0.15, -1.00)\) and a multistable point where PM, FM and antiferromagnetic (AF) solutions coexist: \mkFive~~\(=(-0.05, 0.05)\).
        \textbf{(c)} Largest magnetization value \(m_x = (x_A + x_B)/2\) (see main text) among the stable stationary solutions. Ferromagnetic ordering appears almost exclusively in quadrants II and IV. We plot the largest value of \(m_x\) here, but in the FM regime the stable solutions appear in symmetry-broken pairs \(\pm m_x^{\mathrm{ss}}\). The insets show the stationary values (green stable, red unstable) along the white cuts. (c.1) At \(V=-1\) the FM phase sets in continuously at \(\Omega = 0\), and at \(\Omega \approx 0.15\) the system jumps between two FM states (\mkFFM). (c.2) At \(\Omega = -0.1\) the onset in \(V\) is discontinuous, with stable PM and FM solutions coexisting over a finite range.
        \textbf{(d)} Largest staggered magnetization \(n_x = (x_A - x_B)/2\). The plot is equivalent to (c) when \(V \to -V\) and \(m_x \to n_x\) as a result of the duality transformation (see Sec.~\ref{sec:duality}): antiferromagnetic ordering appears in quadrants I and III, and the insets (d.1) at \(V = +1\) and (d.2) at \(\Omega = -0.1\) correspond to (c.1) and (c.2).}
    \label{fig:phase_diagram}
\end{figure*}

\subsection{Stationary state analysis}
\label{sec:mf_stationary}

\subsubsection{Fixed points}
\label{sec:fixedpoints}
The stationary solutions, or fixed points, of Eqs.~\eqref{eq:sublatticeX}-\eqref{eq:sublatticeZ} obey the stationarity condition \(\dot{\vec m}_A = \dot{\vec m}_B = 0\).  The simplest scenario for solving these stationary equations is the one where the transverse field \(\Omega\) is zero. Here, both sublattices are described by the same Bloch vector, representing a fully polarized state whose Bloch vector is pointing in the negative \(z\)-direction:
\begin{eqnarray}
    \vec m_A=\vec m_B=(0,0,-1).\label{eq:paramagnetic_state}
\end{eqnarray}
This state is completely symmetric, in the sense that it is invariant under both the symmetry transformation \eqref{eq:parity} and the duality transformation \eqref{eq:duality}. At finite transverse field strength \(\Omega \neq 0\), however, the calculations are more involved. Here it is convenient to first express the \(y,z\)-components of the Bloch vectors in terms of the \(x\)-component. Using Eqs.~\eqref{eq:sublatticeX} and \eqref{eq:sublatticeZ} at stationarity, this yields
\begin{equation}
    y_{A/B} = -\frac{\Gamma}{4\Omega}\, x_{A/B},
    \quad
    z_{A/B} = -1 - \frac{h_{AB/BA}\, x_{A/B}}{4\Omega}.
    \label{eq:myz}
\end{equation}
Substituting into Eq.~\eqref{eq:sublatticeY} this eventually leads to the following equations for the \(x\)-component of the Bloch vector on each sublattice
\begin{figure*}[tbp]
    \makebox[\linewidth][r]{
        \includegraphics[width=1.05\linewidth]{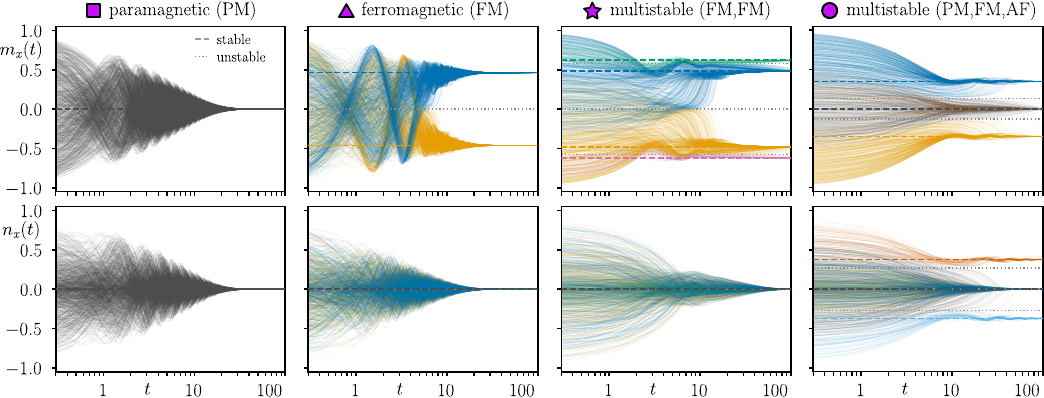}}
    \caption{\textbf{Mean-field dynamical behavior.}
        Initial states relax to the stable stationary solutions analyzed in Fig.~\ref{fig:phase_diagram}(b)-(d). The number of distinct stationary states depends on the parameter regime, which we analyze for the representative points \mkPM, \mkFM, \mkFFM, \mkFive\ in Fig.~\ref{fig:phase_diagram}(b). The top (bottom) row of panels shows the evolution of the magnetization \(m_x\) (staggered magnetization \(n_x\)) (see Sec.~\ref{sec:orderparameters}) on a logarithmic time scale. The trajectories are found by numerical integration of the dynamical mean-field equations \eqref{eq:sublatticeX}-\eqref{eq:sublatticeZ} from 1001 different initial states, sampled as described in Appendix~\ref{app:seeds}. The coloring distinguishes trajectories with different stationary solutions. The colored dashed lines indicate the value of the stationary solutions; the dotted lines correspond to unstable solutions that exist in principle but cannot be reached dynamically (see Sec.~\ref{sec:mf_stability}). Panels (\mkPM): All trajectories decay to the symmetric state with \(m^\mt{PM}_x=n_x^\mt{PM}=0\) [see Eq.~\eqref{eq:paramagnetic_state}] via damped oscillations. Panels (\mkFM): The trajectories relax through damped oscillations towards the two ferromagnetic states with magnetization \(+ m^\mt{FM}_x\) (dark blue) and \(-m^\mt{FM}_x\) (yellow). The corresponding stationary states have \(n_x = 0\) (see Sec.~\ref{sec:fixedpoints}). Panels (\mkFFM): Trajectories relax towards four different ferromagnetic states with magnetization \(\pm m_x^\mt{FM1}\) (dark blue, yellow) and \(\pm m_x^\mt{FM2}\) (green, pink). The two additional solutions are generated by an emergent symmetry (discussed in Appendix~\ref{app:symmetryline}). All the corresponding stationary states have \(n_x = 0\). Panels (\mkFive): The trajectories relax towards five different stationary states, each with a different value for \(m_x\) and \(n_x\). Two ferromagnetic solutions \(\pm m_x^\mt{FM3}\) (dark blue, yellow) and two antiferromagnetic solutions \(\pm n_x^\mt{AF}\) (orange, light blue) coexist with the paramagnetic solution \(m^\mt{PM}_x=n_x^\mt{PM}=0\) (gray).}
    \label{fig:trajectories}
\end{figure*}
\begin{equation}
    \Big(h_{AB/BA}^2 + 8\Omega^2 + \frac{\Gamma^2}{2}\Big)\, x_{A/B} + 4\Omega\, h_{AB/BA} = 0 .
    \label{eq:polynomial}
\end{equation}
Taking the sum and the difference of Eqs.~\eqref{eq:polynomial} yields
\begin{equation}
    \begin{aligned}
        (x_B \pm x_A)\Big[\eta^2 x_A x_B + 4\Omega\eta \pm \Big(8\Omega^2 + \tfrac{\Gamma^2}{2}\Big)\Big] & = 0,
    \end{aligned}
    \label{eq:factorized}
\end{equation}
with \(\eta = 4V + 8 x_A x_B\).
Here it becomes apparent that there are two classes of solutions: one where the sublattices are uniformly occupied and the system is translation invariant \({x_A = x_B}\), and one with staggered occupation \(x_A = - x_B\). In the following we briefly describe the general structure of the possible solutions, leaving aside for the moment considerations of their stability under fluctuations.

\emph{Paramagnetic fixed point.} The trivial solution \({x_A = x_B = 0}\) always exists. As discussed at the beginning of this subsection the associated sublattice Bloch vectors are given by \(\vec m_A = \vec m_B = (0, 0, -1)\), i.e., the state is fully polarized in \(z\)-direction. We refer to this state as paramagnetic because both order parameters are zero: \(m^\mt{PM}_x=n^\mt{PM}_x=0\).

\emph{Ferromagnetic (uniform) fixed points.}
Next, we consider the class of fixed points where the sublattices are occupied equally and feature a non-zero magnetization in \(x\)-direction, i.e. \(x_A = x_B \neq 0\). Here, the magnetization is finite (\(m_x\neq 0\)) and the staggered magnetization is zero (\(n_x = 0\)). In this case the mean-field equations \eqref{eq:factorized} reduce to the cubic polynomial
\begin{eqnarray}
    u^3 + V u^2 + b u + c &=& 0,
    \label{eq:cubic_m}
\end{eqnarray}
which depends on the parameters
\begin{eqnarray*}
    b = \frac{V^2 + 2\Omega}{4},\qquad  c = \frac{16\Omega^2 + 32 V\Omega + \Gamma^2}{128}
\end{eqnarray*}
and is expressed in terms of the variable \(u = m_x^2\), i.e. the square of the \(x\)-magnetization. Physical solutions therefore have to obey \(u\geq 0\). If those exist, they are associated with pairs of stationary solutions whose magnetization is \(m_x = \pm\sqrt{u}\).

\emph{Antiferromagnetic (staggered) fixed points.} Analogously, there are stationary solutions for which the magnetization of the two sublattices is finite but have opposite signs. Here we have  \(x_A = - x_B \neq 0\) and thus \(m_x = 0\) and \(n_x\neq 0\). The resulting mean-field equation reads
\begin{eqnarray}
    v^3 - V v^2 + b v + d &=& 0,
    \label{eq:cubic_n}
\end{eqnarray}
with
\begin{eqnarray*}
    d = \frac{16\Omega^2 - 32 V\Omega + \Gamma^2}{128}
\end{eqnarray*}
and the variable \(v = n_x^2\), which shows that also here only solutions with \(v\geq 0\) are physically permissible. Note that Eq.~\eqref{eq:cubic_n} is related to Eq.~\eqref{eq:cubic_m} by the change \(V \rightarrow -V\), which is a direct consequence of the duality transformation discussed in Sec.~\ref{sec:duality}.

Several of the stationary states discussed above can exist in parallel. However, whether they are ultimately approached depends on their stability. Which one is approached, on the other hand, depends on the initial conditions. Both aspects will be discussed in detail in the following sections. As a starting point, we numerically compute the number of real solutions as a function of \(\Omega\) and \(V\), fixing \(\Gamma=0.3\). Figure~\ref{fig:phase_diagram}(a) shows that a maximum of 9 simultaneous fixed points can be realized for this parameter choice. For comparison, it is important to note that in the absence of four-body interactions (\(W=0\)) at most three fixed points exist \cite{maghrebi2016,Boneberg2025,Overbeck2017}. This shows that indeed the inclusion of such interaction terms augments the structure of the manifold of stationary states.

\subsubsection{Stability analysis}
\label{sec:mf_stability}
In general, not all of the stationary fixed points found in Sec.~\ref{sec:fixedpoints} are stable.
To identify the stable ones we perform in the following a linear stability analysis.
To this end, we linearize the six mean-field equations of motion \eqref{eq:sublatticeX}-\eqref{eq:sublatticeZ} around a given fixed point \(\vec M^* = (x^*_A, y^*_A, z^*_A, x^*_B, y^*_B, z^*_B)\). Within this approximation, small deviations from the fixed point \(\delta\vec{M}= \vec{M}-\vec{M}^*\)
follow the linear equation
\[\delta\dot{\vec M} = \mathcal{J}_\mt{FP}\,\delta\vec M,\]
where \(\mathcal{J}_\mt{FP}\) is the Jacobian evaluated at the examined fixed point. The analysis then amounts to determining the sign of the real part of the six eigenvalues of \(\mathcal{J}_\mt{FP}\). If at least one of them has a positive real part, the fixed point is unstable against small perturbations. We provide the general form of \(\mathcal{J}_\mt{FP}\) in Appendix~\ref{app:mf} and discuss here only the case related to the paramagnetic stationary solution, where \(\vec M^* = (0, 0, -1, 0, 0, -1)\).
In this special case the Jacobian takes on the form
\begin{equation}
    \mathcal{J}_{\mathrm{PM}} =
    \begin{pmatrix}
        -\frac{\Gamma}{2} & -2\Omega          & 0       & 0                 & 0                 & 0       \\[2pt]
        2\Omega           & -\frac{\Gamma}{2} & 0       & 4V                & 0                 & 0       \\[2pt]
        0                 & 0                 & -\Gamma & 0                 & 0                 & 0       \\[2pt]
        0                 & 0                 & 0       & -\frac{\Gamma}{2} & -2\Omega          & 0       \\[2pt]
        4V                & 0                 & 0       & 2\Omega           & -\frac{\Gamma}{2} & 0       \\[2pt]
        0                 & 0                 & 0       & 0                 & 0                 & -\Gamma
    \end{pmatrix},
    \label{eq:jac_pm}
\end{equation}
which is diagonalized analytically. Its eigenvalues are
\begin{equation}
    \lambda_{1,2} = -\Gamma\ ,
    \qquad
    \lambda_{3,4,5,6} = -\frac{\Gamma}{2} \pm \sqrt{-4\Omega\left(\Omega \pm 2V\right)}
    \label{eq:pm_eigenvalues}.
\end{equation}
We find \(\Re(\lambda_{1,2}) < 0\) since \(\Gamma > 0\).
For the paramagnetic fixed point to be stable, we thus need to impose for the remaining eigenvalues the condition \(\Re(\lambda_{3,4,5,6}) < 0\). This yields a critical two-body interaction strength, \(V_\mathrm{c}\), below which the stability of the paramagnetic solution is maintained:
\begin{align}
    |V| < V_\mathrm{c} = \frac{|\Omega|}{2} + \frac{\Gamma^2}{32|\Omega|}.
    \label{eq:pm_stability}
\end{align}

The analysis of the remaining fixed points --- which feature a finite magnetization \(m_x\) and/or staggered magnetization \(n_x\) --- requires numerically finding the eigenvalues of the associated Jacobian. The result is displayed in Fig.~\ref{fig:phase_diagram}(b), which shows the number of stable solutions in the \(\Omega\)--\(V\) plane.

\subsubsection{Stationary state phase diagram}
Next, we compute the values of the order parameters associated with the stable fixed points. In some parameter regimes, several stable stationary solutions coexist. In Fig.~\ref{fig:phase_diagram}(c) and Fig.~\ref{fig:phase_diagram}(d) we show the phase diagram for the stationary solutions with the maximum value for \(m_x\) and \(n_x\), respectively. Note, however, that as a result of the parity symmetry, the (staggered) magnetized solutions occur in pairs \(\pm m_x\) (\(\pm n_x\)). This is illustrated in the insets in Fig.~\ref{fig:phase_diagram}(c.1)-(d.2), where we present cuts in the \(\Omega\)--\(V\) plane, depicting the possible stable and unstable solutions. As can be seen by comparing Figs.~\ref{fig:phase_diagram}(c) and (d), symmetry-broken solutions with finite \(n_x\) are predominantly found in the first (I) and third (III) quadrant of the \(\Omega\)--\(V\) plane. Conversely, symmetry-broken solutions with finite \(m_x\) are predominantly found in quadrants II and IV.
The only regions where solutions with finite magnetization \(m_x\) and solutions with finite staggered magnetization \(n_x\) are simultaneously stable are given by the small cusps in quadrant (III) of Fig.~\ref{fig:phase_diagram}(c) and quadrant (II) of Fig.~\ref{fig:phase_diagram}(d) [cf. Fig.~\ref{fig:phase_diagram}(b)].

For the remainder of this work, we choose four representative parameter regimes, marked in Fig.~\ref{fig:phase_diagram}(b), to analyze the fine structure of the phase diagram and the approach to stationarity. We will refer to them as
\begin{itemize}
    \item[\mkPMal] a paramagnetic regime (PM),
    \item[\mkFMal] a uniformly magnetized, i.e. ferromagnetic, regime (FM),
    \item[\vspace{-0.1em}\mkFFMal] a multistable ferromagnetic regime where two pairs of ferromagnetic solutions coexist (FM, FM),
    \item[\mkFiveal] a multistable regime where paramagnetic, ferromagnetic and staggered magnetized solutions coexist (PM, FM, AF).

\end{itemize}

\subsection{Approach to stationarity and basins of attraction}
\label{sec:dynamicsbasins}

\subsubsection{Dynamics}
Our next goal is to understand which stationary state is dynamically approached when starting from a given initial state.
To this end, we integrate Eqs.~\eqref{eq:sublatticeX}-\eqref{eq:sublatticeZ} numerically for \(1001\) different initializations of the Bloch vectors \(\vec m_A , \vec m_B\) of the two sublattices, and visualize the trajectories of the order parameters \(m_x(t)\) and \(n_x(t)\) in Fig.~\ref{fig:trajectories}. The sampling scheme for the initial configurations is described in detail in Appendix~\ref{app:seeds}.
\begin{figure*}[tbp]
    \makebox[\linewidth][r]{
        \includegraphics[width=1.06\linewidth]{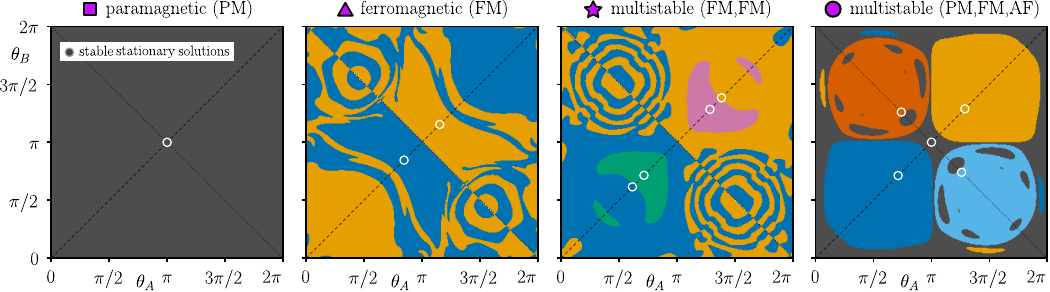}}
    \caption{\textbf{Basins of attraction.} The space of initial states is partitioned into basins, each of which contains all initial states that evolve towards a specific stationary solution [cf. Fig.~\ref{fig:phase_diagram}(b)]. The boundaries that define these basins of attraction are highly nonlinear. We map them out by numerically integrating the mean-field Eqs.~\eqref{eq:sublatticeX}-\eqref{eq:sublatticeZ} for initial conditions \(\vec m_{A/B}(0) = (\sin\theta_{A/B}, 0, \cos\theta_{A/B})\) parametrized by independent angles \(\theta_{A/B}\) on each sublattice. We integrate up to \(t=400\). Initial states with the same stationary state are illustrated in the same color (cf. color coding in Fig.~\ref{fig:trajectories}). White circles mark the projection \(\theta_{A/B} = \operatorname{atan2}(x^*_{A/B}, z^*_{A/B})\) of the stationary Bloch vectors onto the sampling plane. The stationary states themselves [see Eq.~\eqref{eq:meanfieldstate}] can lie inside the Bloch sphere, so they are not themselves initial states of the scan (see Sec.~\ref{sec:fixedpoints}).
    Panel (\mkPM): All initial states relax to the paramagnetic state \(m_x^{\mathrm{PM}} = n_x^{\mathrm{PM}} = 0\) (gray).
    Panel (\mkFM): Two basins (dark blue, yellow), one for each stationary solution \(\pm m_x^{\mathrm{FM}}\). Their boundaries are symmetric about the diagonal/anti-diagonal.
    Panel (\mkFFM): Four basins that correspond to the ferromagnetic solutions \(\pm m_x^{\mathrm{FM1}}\) (dark blue, yellow) and the ferromagnetic solutions \(\pm m_x^{\mathrm{FM2}}\) (green, pink).
    Panel (\mkFive): Five basins, corresponding to the two ferromagnetic solutions \(\pm m_x^{\mathrm{FM3}}\) (dark blue, yellow), the two antiferromagnetic solutions \(\pm n_x^{\mathrm{AF}}\) (orange, light blue) and the paramagnetic solution (gray).
    In all plots we fixed system parameters \(W=1, \Gamma = 0.3\) and sampled 300 angles \(\theta_A\) and \(\theta_B\), respectively, in the interval \([0,2\pi)\).
    }
    \label{fig:basins}
\end{figure*}
As can be seen in Fig.~\ref{fig:trajectories}, in the paramagnetic regime (\mkPM) both magnetizations, \(m_x(t)\) and \(n_x(t)\), indeed approach zero for sufficiently long times \(t\).
In the ferromagnetic regime~(\mkFM) ergodicity is broken. Each trajectory ends at one of two fixed points with finite magnetization, which are related by the parity symmetry (see Section~\ref{sec:parity}). There is no stable fixed point with finite \(n_x\), so the staggered magnetization decays to zero.
In fact this regime also exists in the absence of four-body interactions \(W = 0\), where it has been used to implement a binary classification of the initial states in quantum neural networks~\cite{Boneberg2025}. Due to duality, for \(V \rightarrow -V\) there also exists an inverse (antiferromagnetic) phenomenology where \(m_x\) decays to zero and \(n_x\) remains finite.

In the multistable regime~(\mkFFM) two stable ferromagnetic fixed points coexist, each with its parity-related partner. This gives rise to four stationary states. Hence, the dynamics allows one to separate the set of initial states into four distinct classes, each of which is displayed in a different color in Fig.~\ref{fig:trajectories}. The occurrence of these four ferromagnetic solutions is due to the breaking of an emergent \(\mathbb{Z}_2\) symmetry (see Ref.~\cite{marcuzzi_universal_2014}) in addition to the parity symmetry. We discuss this in detail in Appendix~\ref{app:symmetryline}. Again, due to duality, for \(V \rightarrow -V\) there exists an antiferromagnetic phenomenology where \(m_x\) decays to zero and \(n_x\) remains finite. In the second multistable regime~(\mkFive), the paramagnetic solution and stationary solutions with finite \(m_x\) and \(n_x\) coexist. In total, this gives rise to five distinct stationary states that can be used to classify the set of initial states.

\subsubsection{Basins of attraction}
Now we want to analyze in some more detail which initial states are attracted to which stationary mean-field solution. This will allow us to map out the so-called basins of attraction. To reduce complexity, we restrict the initial states to pure product states whose sublattice Bloch vectors lie in the \(x\)--\(z\) plane. Each is specified by a single angle per sublattice,
\begin{equation}
    \vec{m}_{A/B}(0) = (\sin\theta_{A/B},\, 0,\, \cos\theta_{A/B}),
    \label{eq:init_angles}
\end{equation}
and we scan both angles on a uniform grid of \(300 \times 300\) points in \([0, 2\pi)^2\). In contrast to the mixed-state sampling used for Fig.~\ref{fig:trajectories} (Appendix~\ref{app:seeds}), every initial state here lies on the surface of the Bloch sphere.
For each initial state, specified by the pair \((\theta_A, \theta_B)\), we integrate the mean-field equations \eqref{eq:sublatticeX}-\eqref{eq:sublatticeZ} until a fixed point is reached. To map out the associated basin of attraction we assign a color to each fixed point (in analogy to Fig.~\ref{fig:trajectories}) and mark the point in the \(\theta_A\)--\(\theta_B\) plane with the corresponding color. This yields the plots shown in Fig.~\ref{fig:basins}. Therein, we mark the fixed points of Fig.~\ref{fig:trajectories} as white circles.

Evidently, the paramagnetic regime (\mkPM) is ergodic, i.e., all initial states are drawn to the paramagnetic fixed point.
In the other parameter regimes of interest, ergodicity is broken, but the number of basins of attraction varies between two (\mkFM), four (\mkFFM) and five (\mkFive), in correspondence with the number of stable fixed points that were identified in Fig.~\ref{fig:trajectories}.
In general, the basins form intricate interwoven structures, whose boundaries are highly nonlinear. These structures correspond to complex decompositions of the input space and represent solutions of highly non-trivial classification tasks.

\section{Finite-size numerics}
\label{sec:ed_method}
\begin{figure*}[tbp]
    \makebox[1.00\linewidth][r]{
        \includegraphics[width=1.04\linewidth]{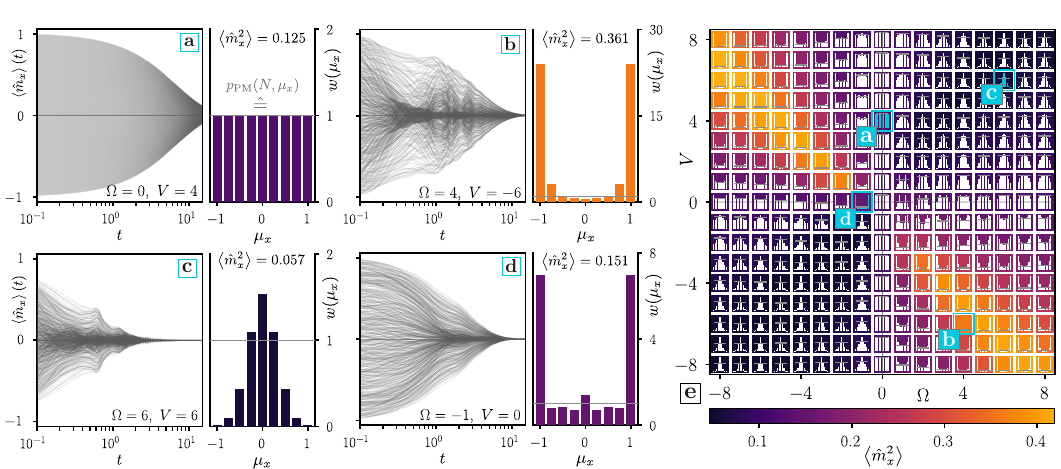}}
    \caption{\textbf{Finite-size numerics.}
        The finite system always relaxes towards \(\ev{\hat{m}_x}=0\), but the stationary states keep signatures of the mean-field phases in the distribution of their magnetization values. Indeed, the expectation value of the second moment of the magnetization operator, \(\hat m_x^2\), reproduces the qualitative structure of the mean-field phase diagram in Fig.~\ref{fig:phase_diagram}. All results are obtained via numerical integration of the master equation \eqref{eq:masterequation}, with periodic boundary conditions, \(N=8\) sites, and system parameters \(W=1, \Gamma=0.3\).
        \textbf{(a)-(d)}
        Left side of panel: 600 evolution trajectories of \(\ev{\hat{m}_x}(t)\) for initial product states uniformly distributed in the interval \(\ev{\hat{m}_x}(0)\in[-1,1]\) on a logarithmic time axis starting at \(t=0.1\). All trajectories approach \(\ev{\hat{m}_x}=0\), in all panels.
        Right side of panel: relative distribution of magnetizations, \(w(\mu_x)\) [see Eq.~\eqref{eq:weight}], in the stationary state. The gray line, \(w=1\), marks the reference value where the absolute distribution of the magnetization [see Eq.~\eqref{eq:distributionfunction}] corresponds to the paramagnetic state, i.e. is given by the binomial distribution \(p_{\mathrm{PM}}(N,\mu_x)\) (see main text).
        (a) \(\Omega=0\), \(V=4\): The Hamiltonian commutes with all \(\sigma_k^x\) and the decay damps all correlations between them. Therefore, the relative distribution is uniform as in the paramagnetic state. The associated second moment of the magnetization distribution is \(\ev{\hat m^2_x} = 1/N= 0.125\).
        (b) \(\Omega=4\), \(V=-6\): the relative distribution features peaks at \(\mu_x=\pm1\). We interpret this bimodality as a finite-size signature of the two ferromagnetic stationary solutions, \(\pm m_x^{\mathrm{FM}}\), appearing in the mean-field study at this parameter choice [cf. Fig.~\ref{fig:phase_diagram}(c)]. This distribution also has a larger second moment \(\ev{\hat m^2_x}=0.361>1/N\).
        (c) \(\Omega=6\), \(V=6\): distribution concentrated around \(\mu_x = 0\), \(\ev{\hat m^2_x}=0.057<1/N\). We interpret this single peak as a signature of the antiferromagnetic stationary solutions found in the mean-field study of this parameter regime [cf. Fig.~\ref{fig:phase_diagram}(d)].
        (d) \(\Omega=-1\), \(V=0\): relative distribution features two large peaks at \(\mu_x=\pm1\) and a smaller peak at \(\mu_x=0\); we interpret this as a signature of the multistable regime, where FM and AF solutions coexist [cf. Fig.~\ref{fig:trajectories}, panel (\mkFive)].
        \textbf{(e)} The second moment \(\ev{\hat m^2_x}\) as a function of \(\Omega\) and \(V\). This color plot is divided into cells which contain a histogram illustrating the shape of the relative distribution in the corresponding parameter region. As indicated by the discussion of the special parameter regions (a-d), the occurrence of histogram modes broadly aligns with the occurrence of stationary solutions in the mean-field study [cf. Fig.~\ref{fig:phase_diagram}(c),(d)].
    }
    \label{fig:distquench}
\end{figure*}
In the following, we investigate whether and how the findings of the mean-field analysis manifest in finite systems.
We integrate the master equation~\eqref{eq:masterequation} for a chain with \(N = 8\) sites.
The corresponding data is shown in Fig.~\ref{fig:distquench}.
We want to discuss this by first focusing on the dynamics, in particular on the evolution of the expectation value of the order parameter \(\hat m_x\).
This is represented in the four panels (a,b,c and d) that display the evolution of \(\ev{\hat{m}_x}\) for different initial states analogous to the mean-field data shown in Fig.~\ref{fig:trajectories}. Each panel refers to a different choice of parameters \(V\) and \(\Omega\).
The first observation is that for long times all initial conditions evolve towards \(\expval{\hat m_x}=0\), regardless of the initial state and parameter choice.
At first glance, this is at odds with the mean-field analysis, but this is merely a consequence of the fact that in a finite system non-ergodic behavior and symmetry breaking do not occur \cite{goldenfeld2018}.

However, as we show next, some signatures of the physics that we observed in the mean-field analysis become indeed visible at the level of the distribution function of the magnetization.
To obtain the distribution function, we make use of the fact that we can write the magnetization as
\begin{eqnarray}
    \hat m_x =\sum_{{\mu_x}} \mu_x Q(\mu_x)
\end{eqnarray}
where  \(\mu_x \in \{-1, -1+\tfrac{2}{N}, \dots, 1-\tfrac{2}{N}, 1\}\) denotes the possible magnetization values and \(Q(\mu_x)\) is the projector onto the space of spin configurations that carry magnetization \(\mu_x\).
The distribution function of magnetization in a state \(\rho\) is thus
\begin{equation}
    p(\mu_x) = \tr[Q(\mu_x)\, \rho].
    \label{eq:distributionfunction}
\end{equation}
For the following analysis we take as a reference the paramagnetic mean-field (product) state with the sublattice Bloch vectors \eqref{eq:paramagnetic_state}, for which the distribution function is the binomial distribution
\begin{eqnarray}
    p_\mathrm{PM}(N,\mu_x)=\frac{1}{2^N}\binom{N}{\frac{N}{2}(\mu_x+1)}.
    \label{eq:binom}
\end{eqnarray}
In order to quantify how the numerically computed distribution function \(p(\mu_x)\) differs from this reference, we study in the following the relative weights
\begin{equation}
    w(\mu_x) = \frac{p(\mu_x)}{p_\mathrm{PM}(N,\mu_x)}.
    \label{eq:weight}
\end{equation}
When the system approaches the paramagnetic state one thus has \(w(\mu_x) = 1\) for all \(\mu_x\).

Let us now focus on the data shown in Fig.~\ref{fig:distquench}(a), which was computed at vanishing transverse field strength, \(\Omega=0\). Here the mean-field solution predicts that the system approaches the paramagnetic state \eqref{eq:paramagnetic_state}, which is the case marked by \mkPM\, in Figs.~\ref{fig:trajectories} and~\ref{fig:basins}. Indeed, also the finite-size simulations confirm this, as the relative weights are \(w(\mu_x)=1\), uniformly. In Fig.~\ref{fig:distquench}(b) we consider a parameter set with finite transverse field strength \(\Omega\) and negative two-body interaction \(V\). The mean-field phase diagram in Fig.~\ref{fig:phase_diagram}(c) predicts here the evolution towards a ferromagnetic stationary state (case \mkFM\, in Fig.~\ref{fig:trajectories}). Indeed, an inspection of the relative weights \(w(\mu_x)\) shows that the probability to find the system in ferromagnetic spin states with the magnetization eigenvalues \(\mu_x = \pm 1\) is enhanced compared to the paramagnetic distribution, which is indicated as gray line. Likewise, we find for the data shown in Fig.~\ref{fig:distquench}(c), with parameters as illustrated in Fig.~\ref{fig:distquench}(e), that at stationarity more relative weight is on spin states with small absolute magnetization. This may be interpreted as a signature of the antiferromagnetic stationary state that the mean-field analysis predicts here [see Fig.~\ref{fig:phase_diagram}(d)]. As the final case, we show in Fig.~\ref{fig:distquench}(d) a situation where states with small and large magnetization carry a relatively high weight. This could be interpreted as a signature of the multistable regime (\mkFive) depicted in Figs.~\ref{fig:trajectories} and~\ref{fig:basins}.

The distribution function of the magnetization becomes wider (narrower) compared to the paramagnet, when the system approaches a ferromagnetic (antiferromagnetic) stationary state. Given that the average magnetization is always zero in the considered finite system this change in width is captured by the second moment \(\ev{\hat m_x^2}\), which we show in Fig.~\ref{fig:distquench}(e), along with the relative weights \eqref{eq:weight}, over a broad range of parameters. Qualitatively, the structure of this plot coincides with that of the mean-field phase diagram depicted in Fig.~\ref{fig:phase_diagram}(c): the existence of multiple fixed points in the mean-field study at a certain point in parameter space seems to manifest in a multimodal distribution of the order parameter at finite size. This may be seen as a hint that the mean-field study provides a meaningful description of the infinite-size stationary state manifold.

\section{Conclusions and outlook}
\label{sec:conclusion}
We have investigated the dynamics of a dissipative quantum Ising model with competing two- and four-body interactions. This was motivated by the fact that this and related models can effectively describe the processing of data in certain quantum neural network architectures \cite{gillman2022b,Boneberg2023,BonebergNonlinear2025,Boneberg2025}. In particular, the capability of these networks to classify data can be pinned down to the emergence of multiple metastable states \cite{Boneberg2025,BonebergNonlinear2025} that are transient attractors of the dynamics. The character of such metastable states can often be assessed through a mean-field analysis of the system. In this regard, our work sheds light on whether and how metastable states --- in particular their number --- could be controlled by the introduction of multi-body interactions which would correspond to multi-body gates in the associated quantum neural network. We showed that their incorporation indeed enables multi-class classification manifesting in the presence of up to five stable stationary states for the specific model at hand. Moreover, the intricate shape of the associated basins of attraction hints towards the ability to realize complex and nonlinear decision boundaries.

A limiting aspect of our investigation is certainly that the majority of its findings are based on a mean-field analysis. Small-scale numerical calculations provided qualitative evidence for the applicability of the conclusions drawn from mean-field. However, a large-scale numerical study, e.g., based on tensor network algorithms, is certainly in order. If these studies are affirmative, multi-body gates and the associated many-body interactions may indeed turn out to be a resource to control the expressivity of quantum neural networks. For example, increasing the order of the interactions may increase the number of classes that can be distinguished. That this is a reasonable assumption is visible at the level of the mean-field equations \eqref{eq:polynomial}. Here higher order interactions increase the degree of the involved polynomials and the ensuing mean-field equation will generally acquire more and more stationary solutions.

\textit{Data availability.}---The data and code that support the findings of this article are openly available on Zenodo~\cite{KochsiekZenodo2026}.

\textit{AI usage.}---The authors acknowledge using Anthropic's Claude Opus~5.5 for language editing and for formatting the \LaTeX{} source, and Claude Opus~4.8 for refactoring the authors' code for publication. All suggestions were reviewed and approved by the authors.
\begin{acknowledgments}
    We acknowledge funding from the Deutsche Forschungsgemeinschaft (DFG, German Research Foundation) through the Research Unit FOR 5413 and support from the Leverhulme Trust (Grant No. RPG-2024-112). This work is supported through the Federal Ministry of Research, Technology and Space (BMFTR) via the project NeuQuant (Grant No. 13N17065). IL is a member of the Machine Learning Cluster of Excellence, funded by the Deutsche Forschungsgemeinschaft (DFG, German Research Foundation) under Germany’s Excellence Strategy—EXC Number 2064/1 - Project Number 390727645. This work is supported by ERC grant OPEN-2QS (Grant No. 101164443). The authors acknowledge support by the state of Baden-W\"urttemberg through bwHPC and the German Research Foundation (DFG) through Grant No. INST 40/575-1 FUGG (JUSTUS 2 cluster).
\end{acknowledgments}

\bibliographystyle{apsrev4-2}
\bibliography{references}
\clearpage

\appendix

\section{Mean-field equations and their linearization}
\label{app:mf}

\subsection{Exact equations for a single site}
We apply the adjoint master equation \eqref{eq:adjoint} to the Pauli operators \(\sigma_k^\alpha\) of a single site \(k\). Every interaction term of the Hamiltonian \eqref{eq:hamiltonian} that contains site \(k\) acts on it through the product of the \(\sigma_j^x\) on its other sites \(j\), which lie at most three sites away. We collect all of these multi-site products in a single operator
\begin{equation}
    \hat h_k = 2V\sum_{j\,\mathrm{nn}\,k}\sigma_j^x + 2\sum_{C\ni k}\,\prod_{j\in C\setminus\{k\}}\sigma_j^x ,
    \label{eq:fieldoperator}
\end{equation}
where the first sum runs over the two nearest neighbors of \(k\) and the second over the four quartets \(C\) of consecutive sites that contain \(k\). The factors of two stem from the commutator \(\commutator{\sigma^x}{\sigma^y} = 2i\sigma^z\). With the jump operators \eqref{eq:jumpoperators} we find,
\begin{align}
    \partial_t\expval{\sigma_k^x} & = -2\Omega\expval{\sigma_k^y} - \frac{\Gamma}{2}\expval{\sigma_k^x}, \nonumber                                         \\
    \partial_t\expval{\sigma_k^y} & = 2\Omega\expval{\sigma_k^x} - \expval{\hat h_k\,\sigma_k^z} - \frac{\Gamma}{2}\expval{\sigma_k^y}, \label{eq:persite} \\
    \partial_t\expval{\sigma_k^z} & = \expval{\hat h_k\,\sigma_k^y} - \Gamma\left(\expval{\sigma_k^z} + 1\right). \nonumber
\end{align}
The interactions do not enter the first line because they commute with \(\sigma_k^x\). The equations are not closed: the expectation values of single spins couple to two- and four-point correlation functions found in the operator \(\hat{h}_k\).

\subsection{Two-sublattice mean-field equations}
We now insert the product ansatz of Sec.~\ref{sec:mf_sublattice}, which is parametrized by the two sublattice Bloch vectors of Eq.~\eqref{eq:blochvectors}. This ansatz factorizes every expectation value of operators on different sites. Consider a site \(k\) on sublattice \(A\). Both of its nearest neighbors belong to \(B\). Each of the four quartets that contain \(k\) consists of two \(A\) and two \(B\) sites, so besides \(k\) it contains one \(A\) and two \(B\) sites. Hence
\begin{equation}
    \expval{\hat h_k\,\sigma_k^\nu} = \left(2\cdot 2V\,x_B + 2\cdot 4\, x_A x_B^2\right)\nu_A = h_{AB}\,\nu_A
\end{equation}
for \(\nu = y, z\), with the effective field \(h_{AB}\) of Eq.~\eqref{eq:fields}. Inserting this into Eq.~\eqref{eq:persite} gives Eqs.~\eqref{eq:sublatticeX}-\eqref{eq:sublatticeZ}. For a site on \(B\) the roles of the sublattices are exchanged, and the field is \(h_{BA}\).

We discuss the following two cases explicitly, because they become relevant in Sec.~\ref{sec:mf_stationary}.

\emph{Uniform states.} If both sublattices have the same Bloch vector, \(\vec m_A = \vec m_B\) with \(x_A = x_B = m_x\), the effective fields coincide,
\begin{equation}
    h_{AB} = h_{BA} = h = 4V m_x + 8 m_x^3 .
    \label{eq:homfield}
\end{equation}
The equations for \(A\) and \(B\) are then identical, and the six equations reduce to three. This is the case of the uniform fixed points, Eq.~\eqref{eq:cubic_m}.

\emph{Staggered states.} If the in-plane components have opposite signs, \(x_A = -x_B = n_x\), \(y_A = -y_B\), \(z_A = z_B\), one finds \(h_{BA} = -h_{AB}\) with \(h_{AB} = -4V n_x + 8 n_x^3\). The equations for \(B\) are then the negative of those for \(A\) in the \(x\) and \(y\) components, and again three equations remain. They are those of the uniform case with \(V \to -V\) and \(m_x \to n_x\), which is the duality of Sec.~\ref{sec:duality} at the mean-field level and leads to Eq.~\eqref{eq:cubic_n}.

\subsection{Jacobian}
To linearize Eqs.~\eqref{eq:sublatticeX}-\eqref{eq:sublatticeZ} around a fixed point \(\vec M^* = (x^*_A, y^*_A, z^*_A, x^*_B, y^*_B, z^*_B)\) we need the derivatives of the fields,
\begin{equation}
    \begin{gathered}
        \pdv{h_{AB}}{x_A} = 8x_B^2, \qquad \pdv{h_{BA}}{x_B} = 8x_A^2, \\
        \pdv{h_{AB}}{x_B} = \pdv{h_{BA}}{x_A} = 4V + 16x_A x_B \equiv \kappa .
    \end{gathered}
    \label{eq:kappa}
\end{equation}
With \(\delta\vec M = (\delta x_A, \delta y_A, \delta z_A, \delta x_B, \delta y_B, \delta z_B)\) the Jacobian reads
\begin{widetext}
    \begin{equation}
        \mathcal{J}_\mt{FP} =
        \begin{pmatrix}
            -\frac{\Gamma}{2}       & -2\Omega          & 0       & 0                       & 0                 & 0       \\[2pt]
            2\Omega - 8 x_B^2\, z_A & -\frac{\Gamma}{2} & -h_{AB} & -\kappa\, z_A           & 0                 & 0       \\[2pt]
            8 x_B^2\, y_A           & h_{AB}            & -\Gamma & \kappa\, y_A            & 0                 & 0       \\[2pt]
            0                       & 0                 & 0       & -\frac{\Gamma}{2}       & -2\Omega          & 0       \\[2pt]
            -\kappa\, z_B           & 0                 & 0       & 2\Omega - 8 x_A^2\, z_B & -\frac{\Gamma}{2} & -h_{BA} \\[2pt]
            \kappa\, y_B            & 0                 & 0       & 8 x_A^2\, y_B           & h_{BA}            & -\Gamma
        \end{pmatrix},
        \label{eq:jac_general}
    \end{equation}
\end{widetext}
where all entries are evaluated at \(\vec M^*\). We diagonalize this matrix numerically at every fixed point.

\emph{Uniform fixed points.} At a uniform fixed point, \(x_A = x_B = m_x\), \(y_A = y_B = y\), \(z_A = z_B = z\), the Jacobian is unchanged under the exchange \(A \leftrightarrow B\). It therefore splits into two \(3\times3\) blocks, one for uniform perturbations (\(\delta\vec m_A = \delta\vec m_B\)) and one for staggered perturbations (\(\delta\vec m_A = -\delta\vec m_B\)),
\begin{equation}
    \mathcal{J}_\pm =
    \begin{pmatrix}
        -\frac{\Gamma}{2}    & -2\Omega          & 0       \\[2pt]
        2\Omega - h'_\pm\, z & -\frac{\Gamma}{2} & -h      \\[2pt]
        h'_\pm\, y           & h                 & -\Gamma
    \end{pmatrix},
    \label{eq:jac_blocks}
\end{equation}
with \(h\) from Eq.~\eqref{eq:homfield} and
\begin{equation}
    \begin{aligned}
        h'_+ & = 8m_x^2 + \kappa = 4V + 24 m_x^2,              \\
        h'_- & = 8m_x^2 - \kappa = -\left(4V + 8 m_x^2\right).
    \end{aligned}
    \label{eq:hprime}
\end{equation}
The block \(\mathcal{J}_+\) is the Jacobian of the three uniform equations, since \(h'_+ = \partial h/\partial m_x\). The block \(\mathcal{J}_-\) describes the stability of a uniform state against staggered perturbations.

\emph{Paramagnetic fixed point.} At \(\vec M^* = (0, 0, -1, 0, 0, -1)\) we have \(m_x = y = h = 0\), \(z = -1\) and \(h'_\pm = \pm 4V\). The full Jacobian becomes Eq.~\eqref{eq:jac_pm}, and the blocks reduce to
\begin{equation}
    \mathcal{J}_\pm^\mt{PM} =
    \begin{pmatrix}
        -\frac{\Gamma}{2} & -2\Omega          & 0       \\[2pt]
        2\Omega \pm 4V    & -\frac{\Gamma}{2} & 0       \\[2pt]
        0                 & 0                 & -\Gamma
    \end{pmatrix}.
\end{equation}
Each block has the eigenvalue \(-\Gamma\) and the pair \(-\Gamma/2 \pm \sqrt{-4\Omega(\Omega \pm 2V)}\), where the inner sign is \(+\) for \(\mathcal{J}_+\) and \(-\) for \(\mathcal{J}_-\). These are the eigenvalues \eqref{eq:pm_eigenvalues}. Uniform perturbations destabilize the paramagnet for \(\Omega V < 0\), staggered ones for \(\Omega V > 0\).

\emph{Staggered fixed points.} Flipping the in-plane components on sublattice \(B\), \((x_B, y_B) \to (-x_B, -y_B)\), and at the same time \(V \to -V\) leaves Eqs.~\eqref{eq:sublatticeX}-\eqref{eq:sublatticeZ} invariant. This maps every staggered fixed point onto a uniform fixed point at \(-V\), and the two Jacobians are related by a change of sign of the corresponding rows and columns. They have the same eigenvalues, so a staggered fixed point at \(V\) is stable if and only if its uniform partner at \(-V\) is.

\section{Initial states of the mean-field trajectories}
\label{app:seeds}
For the mean-field trajectories in Fig.~\ref{fig:trajectories} we construct 1001 initial product states that cover the full range of the magnetization \(m_x\), while the remaining components of the two sublattice Bloch vectors are drawn at random. For each state \(j = 0, \dots, 1000\):
\begin{enumerate}
    \item Magnetization from the grid
          \[
              m_x^{(j)} = -1 + \frac{2j}{1000}
          \]
    \item Staggered magnetization, random
          \[
              n_x^{(j)} \sim \mathcal{U}\big[-(1-|m_x^{(j)}|),\; 1-|m_x^{(j)}|\big]
          \]
    \item \(x\)-components from \(m_x\) and \(n_x\). Inverting
          \(m_x = (x_A + x_B)/2\) and \(n_x = (x_A - x_B)/2\) gives
          \[
              x_A = m_x + n_x, \qquad x_B = m_x - n_x .
          \]
          The bound on \(n_x\) in step 2 guarantees \({|x_A|, |x_B| \le 1}\).
    \item \(y\) and \(z\) for each sublattice \(s = A, B\), random. The Bloch vector must
          satisfy \(x_s^2 + y_s^2 + z_s^2 \le 1\), so \((y_s, z_s)\) must lie in a disc
          of radius \(r_\text{max} = \sqrt{1 - x_s^2}\). A point is drawn uniformly
          over the area of that disc:
          \[
              \xi \sim \mathcal{U}[0,1], \quad \varphi \sim \mathcal{U}[0, 2\pi), \quad
              r = r_\text{max}\sqrt{\xi},
          \]
          \[
              y_s = r\cos\varphi, \qquad z_s = r\sin\varphi .
          \]
          This is done independently for \(s = A\) and \(s = B\), with separate \(\xi\)
          and \(\varphi\). The factor \(\sqrt{\xi}\) is needed for a uniform area
          density. Drawing \(r\) uniformly would put too many points near the center,
          because the area of a ring grows with \(r\).
    \item Resulting distribution of \(n_x\). Since \(n_x\) is drawn conditionally on
          \(m_x\), its marginal density is not flat. Averaging the uniform distribution
          of step 2 over the grid in \(m_x\) gives
          \begin{align*}
              f(n_x) & = \int_{|m_x| \le 1-|n_x|} \frac{dm_x}{2}\frac{1}{2(1-|m_x|)} \\
                     & = -\frac{1}{2}\ln|n_x|,\qquad -1 \le n_x \le 1.
          \end{align*}
          It diverges logarithmically at \(n_x = 0\) and vanishes at \(|n_x| = 1\), so most
          initial states have small \(|n_x|\). The two endpoints \(m_x = \pm 1\) are
          deterministic, since they force \(n_x = 0\) and \(y_s = z_s = 0\).
\end{enumerate}

\begin{figure*}[tbp]
    \centering
    \includegraphics[width=\textwidth]{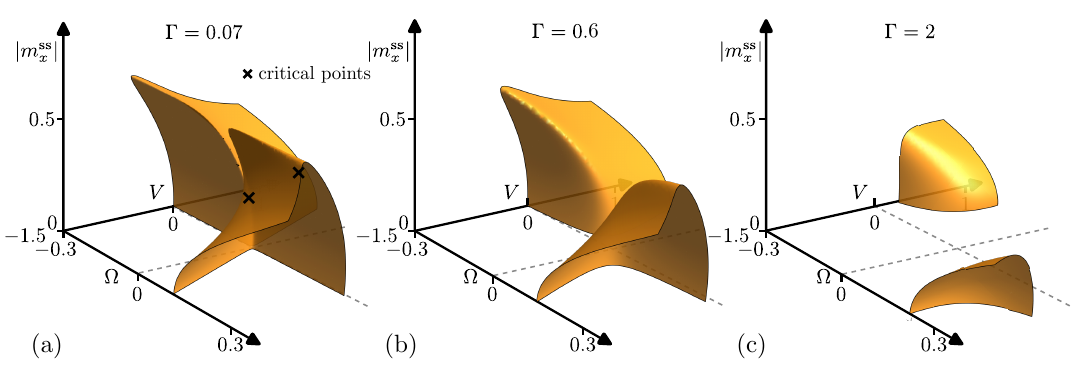}
    \caption{\textbf{Solution surfaces of the cubic polynomial \eqref{eq:cubic_m}.}
        Magnetization \(|m_x^{\mathrm{ss}}| = \sqrt{u}\) of all uniform fixed points, i.e., the non-negative roots \(u\) of Eq.~\eqref{eq:cubic_m}, over the \((\Omega, V)\) plane for \(\Gamma = 0.07\), \(0.6\) and \(2\). Stability is not taken into account. \textbf{(a)} For \(\Gamma < \Gamma_c^{\max}\) the surface at \(\Omega > 0\) folds over twice between the two critical points of Eq.~\eqref{eq:critline} and has three sheets [(\mkFFM) region]. The crosses mark the two critical points. At \(\Omega < 0\) it folds once (coexistence of PM and FM solutions). \textbf{(b)} For \(\Gamma > \Gamma_c^{\max} = 2/(3\sqrt{3}) \approx 0.385\) the critical points and the three-sheeted region are gone, the single fold at \(\Omega < 0\) remains. \textbf{(c)} For strong dissipation magnetized solutions exist only at larger \(|\Omega|\) and \(|V|\). Staggered fixed points follow from \(V \to -V\).}
    \label{fig:solutionsurfaces}
\end{figure*}

\section{Critical manifold and emergent symmetry}
\label{app:symmetryline}
At the point \mkFFM\ the mean-field dynamics has four stable ferromagnetic stationary states, two pairs \(\pm m_x^\mt{FM1}\) and \(\pm m_x^\mt{FM2}\) (Sec.~\ref{sec:dynamicsbasins}). The parity symmetry of Sec.~\ref{sec:parity} is the reason for the fact that every solution comes with a partner of opposite sign. In the \mkFFM\ region there are two pairs of different magnitude. In this appendix we show that the second pair results from a further \(\mathbb{Z}_2\) symmetry. It is not a symmetry of the model, but emerges in the mean-field equations close to a critical point.

\emph{Critical manifold.} The uniform fixed points are given by the non-negative roots \(u = m_x^2\) of the cubic \eqref{eq:cubic_m}. The number of roots changes where two of them merge, the critical manifold is defined as the subspace of system parameters where all three roots coincide, \((u - u_c)^3 = 0\). Comparing the coefficients with Eq.~\eqref{eq:cubic_m}, \(V = -3u_c\), \(b = 3u_c^2\) and \(c = -u_c^3\), gives
\begin{equation}
    u_c = -\frac{V}{3},
    \quad
    \Omega_c = \frac{V^2}{6},
    \quad
    \Gamma_c = \frac{2\sqrt{-V^3\left(4 + 3V\right)}}{3\sqrt{3}}.
    \label{eq:critline}
\end{equation}
For each \(V\) this is one point in the \(\Omega\)--\(\Gamma\) plane, so Eq.~\eqref{eq:critline} describes a line of critical points. It is physical only for \(-4/3 \le V < 0\), where \(u_c > 0\) and \(\Gamma_c\) is real. The largest value, \(\Gamma_c^{\max} = 2/(3\sqrt{3}) \approx 0.385\), is reached at \(V = -1\) and \(\Omega_c = 1/6\). The decay rate \(\Gamma = 0.3\) used in the main text lies below this value.

\emph{Solution surfaces.} The main text works in the \(\Omega\)--\(V\) plane at fixed \(\Gamma\). There the critical manifold shows up as the points with \(\Gamma_c(V) = \Gamma\). For \(\Gamma < \Gamma_c^{\max}\) this condition has two solutions, so the \((\Omega, V)\) plane contains two critical points. Figure~\ref{fig:solutionsurfaces} shows how the uniform fixed points are organized around them. It plots \(|m_x| = \sqrt{u}\) for all non-negative roots of Eq.~\eqref{eq:cubic_m}, regardless of their stability. The staggered fixed points follow from \(V \to -V\) and \(m_x \to n_x\). For \(\Gamma = 0.07\) [Fig.~\ref{fig:solutionsurfaces}(a)] the critical points lie at \((\Omega, V) \approx (0.008, -0.21)\) and \((0.29, -1.33)\). Between them, at \(\Omega > 0\), the surface folds over twice and has three sheets. This is the (FM,FM) region of the main text. At \(\Omega < 0\) the surface folds only once, and its two sheets correspond to the coexistence of PM and FM solutions seen in the cut of Fig.~\ref{fig:phase_diagram}(c.2). For the value \(\Gamma = 0.3\) of the main text the critical points sit at \((0.076, -0.675)\) and \((0.249, -1.223)\), and the three-sheeted region shrinks to the narrow band that the cut \(V = -1\) in Fig.~\ref{fig:phase_diagram}(c.1) crosses at \(\Omega \approx 0.15\). For \(\Gamma > \Gamma_c^{\max}\) there is no critical point and the three-sheeted region is gone [Fig.~\ref{fig:solutionsurfaces}(b)], while the single fold at \(\Omega < 0\) survives. For strong dissipation [Fig.~\ref{fig:solutionsurfaces}(c)] magnetized solutions exist only at larger \(|\Omega|\) and \(|V|\), in line with the widening of the stable PM region, Eq.~\eqref{eq:pm_stability}.

\emph{Expansion around the critical manifold.} At fixed \(V\) we write \(u = u_c + \tilde u\), \(\Omega = \Omega_c + \delta\Omega\) and \(\Gamma = \Gamma_c + \delta\Gamma\). Since \(3u_c + V = 0\) and \(u_c\) does not depend on \(\Omega\) and \(\Gamma\), the quadratic term of the cubic drops out for all values of the perturbations, and we obtain
\begin{equation}
    \tilde u^3 + p\,\tilde u + q = 0,
    \qquad
    p = \frac{\delta\Omega}{2},
    \label{eq:depressedcubic}
\end{equation}
where \(p\) is exact and, to linear order in the perturbations,
\begin{equation}
    q = \frac{V\left(V + 2\right)}{24}\, \delta\Omega
    + \frac{\sqrt{-V^3\left(4 + 3V\right)}}{96\sqrt{3}}\, \delta\Gamma .
    \label{eq:qlinear}
\end{equation}

\emph{Emergent symmetry.} For \(q = 0\), Eq.~\eqref{eq:depressedcubic} is invariant under \(\tilde u \to -\tilde u\). This is the emergent \(\mathbb{Z}_2\) symmetry. It relates a solution with \(m_x^2\) above \(u_c\) to one below \(u_c\), i.e., two solutions of different magnitude, and is therefore distinct from the parity, which flips the sign of \(m_x\). To linear order, \(q = 0\) holds along the direction
\begin{equation}
    \delta\Gamma = -\frac{4\sqrt{3}\, V\left(V + 2\right)}{\sqrt{-V^3\left(4 + 3V\right)}}\, \delta\Omega ,
    \label{eq:symmetryline}
\end{equation}
which we call the symmetry line. Along it the cubic factorizes, \(\tilde u\,(\tilde u^2 + \delta\Omega/2) = 0\). For \(\delta\Omega > 0\) only \(\tilde u = 0\) is real. For \(\delta\Omega < 0\) two further roots \(\tilde u = \pm\sqrt{-\delta\Omega/2}\) split off continuously. This is the structure of the mean-field Ising model~\cite{marcuzzi_universal_2014}, with \(p\) in the role of the reduced temperature and \(q\) in the role of the magnetic field. The stability analysis of Sec.~\ref{sec:mf_stability} shows that the middle root is unstable and the two outer ones are stable [cf. the cut in Fig.~\ref{fig:phase_diagram}(c.1)]. Each of the outer roots gives a parity pair, so that four stable states result,
\begin{equation}
    \pm m_x^\mt{FM1,FM2} = \pm\sqrt{u_c \mp \sqrt{-\delta\Omega/2}}.
\end{equation}
Together with the unstable pair \(\pm\sqrt{u_c}\) and the paramagnet these are seven fixed points. A trajectory that ends in one of the four states breaks both the parity and the emergent symmetry.

\emph{Away from the symmetry line.} For \(q \neq 0\) the emergent symmetry is explicitly broken, as in an Ising magnet in a magnetic field. The cubic keeps three real roots as long as its discriminant is negative, \(4p^3 + 27q^2 < 0\), i.e., for \(\delta\Omega < 0\) and \(|q| < 2(-\delta\Omega/6)^{3/2}\). This defines a narrow band around the symmetry line. It is bounded by two fold lines, on which the middle root merges with one of the outer roots, and the fold lines meet in a cusp on the critical manifold. In Fig.~\ref{fig:solutionsurfaces} these are the edges of the three-sheeted region. Crossing a fold line removes one of the stable pairs, which causes the jump between the two ferromagnetic states in Fig.~\ref{fig:phase_diagram}(c.1). Inside the band the two stable pairs have different distances from \(u_c\), so the emergent symmetry holds only approximately. The point \mkFFM\ lies inside this band. At \((\Omega, V) = (0.15, -1)\) and \(\Gamma = 0.3\) the cubic has the roots \(u = 0.229\), \(0.375\) and \(0.396\) around \(u_c = 1/3\), and the two stable pairs are \(m_x = \pm 0.48\) and \(m_x = \pm 0.63\).

Using the duality transformation of Sec.~\ref{sec:duality} the same structure exists for the staggered solutions at \(V > 0\), with two antiferromagnetic pairs [Fig.~\ref{fig:phase_diagram}(d.1)].

\emph{Critical exponents.} Close to the critical manifold the solutions therefore behave as in the mean-field Ising model. Along the symmetry line, \(|\tilde u| \sim |\delta\Omega|^{1/2}\), i.e., \(\tilde\beta = 1/2\). Perpendicular to it, at \(\delta\Omega = 0\), the linear term vanishes and \(\tilde u^3 = -q\), so \(|\tilde u| \sim |\delta\Gamma|^{1/3}\), i.e., \(\tilde\delta = 3\).
\end{document}